\pdfoutput=1

\RequirePackage{fix-cm}

\documentclass[twocolumn]{svjour3}

\smartqed

\usepackage{graphicx}
\usepackage{amsmath,amssymb}
\usepackage{txfonts}
\usepackage[numbers,sort&compress]{natbib}
\usepackage{hyperref}

\journalname{Nonlinear Dynamics}

\begin{document}

\title{Elucidating the escape dynamics of the four hill potential}

\author{Euaggelos E. Zotos}

\institute{Department of Physics, School of Science, \\
Aristotle University of Thessaloniki, \\
GR-541 24, Thessaloniki, Greece \\
Corresponding author's email: {evzotos@physics.auth.gr}}

\date{Received: 10 October 2016 / Accepted: 20 February 2017 / Published online: 4 March 2017}

\titlerunning{Elucidating the escape dynamics of the four hill potential}

\authorrunning{Euaggelos E. Zotos}

\maketitle

\begin{abstract}

The escape mechanism of the four hill potential is explored. A thorough numerical investigation takes place in several types of two-dimensional planes and also in a three-dimensional subspace of the entire four-dimensional phase space in order to distinguish between bounded (ordered and chaotic) and escaping orbits. The determination of the location of the basins of escape toward the different escape channels and their correlations with the corresponding escape time of the orbits is undoubtedly an issue of paramount importance. It was found that in all examined cases all initial conditions correspond to escaping orbits, while there is no numerical indication of stable bounded motion, apart from some isolated unstable periodic orbits. Furthermore, we monitor how the fractality evolves when the total orbital energy varies. The larger escape periods have been measured for orbits with initial conditions in the fractal basin boundaries, while the lowest escape rates belong to orbits with initial conditions inside the basins of escape. We hope that our numerical analysis will be useful for a further understanding of the escape dynamics of orbits in open Hamiltonian systems with two degrees of freedom.


\end{abstract}

\section{Introduction}
\label{intro}

Undoubtedly one of the most interesting topics in nonlinear dynamics is the issue of escaping particles in Hamiltonian systems (e.g., \cite{C90,CK92,CKK93,STN02}). Hamiltonian systems with escapes are also known as ``open" Hamiltonian systems, in which there is a finite energy of escape. When the value of the energy is higher than the energy of escape, the equipotential surfaces open and escape channels emerge through which the test particles are free to escape to infinity. It should be emphasized that if a test particle has energy larger than the energy of escape, this does not necessarily mean that it will certainly escape from the system and even if escape does occur, the time required for the escape to occur may be very long compared with the natural crossing time. The literature is replete with works on the field of Hamiltonian systems with escapes (e.g., \cite{BBS09,EP14,KSCD99,NH01,Z14a,Z14b,Z15b,Z16a}).

Nevertheless, the issue of escaping orbits in open Hamiltonian systems is by far less explored than the closely related problem of chaotic scattering. The viewpoint of chaos theory has been used in order to investigate and therefore interpret the phenomenon of chaotic scattering (e.g., \cite{BTS96,BST98,BOG89,CHLG12,DJT12,DGJT14,DJ16,GDJ14,JLS99,JMS95,JMSZ10,JT91,SASL06,SSL07,SS08}). At this point, we would like to point out that all the above-mentioned references regarding the problems of open Hamiltonian systems and chaotic scattering are exemplary rather than exhaustive.

In real scattering events we prepare an initial condition in the incoming asymptotic region and let this orbit run until it ends in the outgoing asymptotic region. However, this is not the situation we have in many escape problems of celestial mechanics. Usually in astronomy the particles under consideration are either created in the interior region (interaction region) and/or they are brought to energies above the escape threshold by interactions with other particles in the interaction region. Therefore we have half of a scattering orbit only. However, if we like we can imagine also the past continuation of the escaping orbit and then we have the full analogy to real scattering events. In this sense escape dynamics can be considered a special case of scattering dynamics.

The scattering orbits themselves are not chaotic. In the past and also in the future limit they converge to simple asymptotic motion. However, there can be chaotic invariant sets (chaotic saddles) also for energies above the escape threshold. And the whole bundle of non-chaotic true scattering orbit flows between this chaotic localized set and casts a kind of shadow image of the localized chaotic set into the asymptotic region (e.g., \cite{JS87}). In this sense the scattering data obtained in the asymptotic region contain the information on the localized chaotic set (e.g., \cite{JT91}). This is the first step for the inverse chaotic scattering problem (e.g., \cite{JLS99}). In fact, this is the essential idea of chaotic scattering and chaotic scattering can therefore also be interpreted as transient chaos (see chapter 6 in the book \cite{LT11}).

The chaotic invariant set contains some outermost elements which act as transition states, i.e. as points of no return. When an orbit crosses one of these transition states from the inside to the outside, then this orbit goes to the outgoing asymptotic region and never returns. In generic cases these transition states sit over the index-1 saddle points of the potential (for good information on transition states and the phase space geometry near such points see \cite{WBW04,WBW05}). For 2-dof systems these transition states are just the Lyapunov orbits over the saddle, while for more degrees of freedom they are normally hyperbolic invariant manifolds (NHIMs).

In Hamiltonian systems with escapes an issue of significant importance is the determination of the basins of escape, similar to the basins of attraction in dissipative systems or even to the Newton-Raphson attracting domains (e.g., \cite{CK07,K08,KGK12,KK14,Z16b}). An escape basin is defined as a local set of initial conditions of orbits for which the test particles escape through a certain exit in the equipotential surface. The basins of escape have been studied in many earlier papers (e.g., \cite{AVS01,AVS09,BBS08,BBS09,Z15a,Z15b}). The reader can find more details regarding basins of escape in \cite{C02}. The boundaries between the basins of escape may be fractal (e.g., \cite{AVS09,BGOB88,dML99,SO00}) or even respect the more restrictive Wada property (e.g., \cite{AVS01,KY91,PCOG96}), in the case where three or more escape channels coexist.

Now one interesting question arises: What happens if the system has escape channels without index-1 saddles, for example when the bottom of the escape channels run at a constant energy? One extreme example for this pathological behaviour is the version of the four hill potential (e.g., \cite{BGO90,SS13}). Our aim here is to learn which properties of this pathological
example coincide with the properties of generic systems, such as the H\'{e}non-Heiles system, and which properties are exceptional. In parallel, it is instructive to see which standard numerical methods for the investigation of escape dynamics still work well and which methods break down. In this paper we shall investigate the escape mechanism of orbits of the four hill potential by applying the same computational methods used in earlier similar papers (e.g., \cite{Z15a,Z15b,Z16a}).

The present paper is organised as follows: In Section \ref{mod} we present the properties of the Hamiltonian system. All the computational methods we used in order to explore the escape dynamics of the orbits are described in Section \ref{cometh}. In the following Section, we conduct a thorough and systematic numerical investigation revealing the escape mechanism of the four hill potential. Our paper ends with Section \ref{disc} where the discussion of our research is given.

\section{Properties of the Hamiltonian system}
\label{mod}

In the following we shall explore the escape properties of the four hill potential
\begin{equation}
V(x,y) = x^2y^2 e^{-\left( x^2 + y^2 \right)}.
\label{phill}
\end{equation}

The equations of motion governing the motion of a test particle with a unit mass $(m = 1)$ are
\begin{align}
\ddot{x} &= - \frac{\partial V}{\partial x} = 2 x y^2 \left(x^2 - 1 \right) e^{-\left( x^2 + y^2 \right)}, \nonumber\\
\ddot{y} &= - \frac{\partial V}{\partial y} = 2 x^2 y \left(y^2 - 1 \right) e^{-\left( x^2 + y^2 \right)},
\label{eqmot}
\end{align}
where, as usual, the dot indicates derivative with respect to the time.

Furthermore, the variational equations, needed for the computation of standard chaos indicators (the SALI in our case, as better explained in the following section), are given by
\begin{align}
\dot{(\delta x)} &= \delta \dot{x}, \nonumber\\
\dot{(\delta y)} &= \delta \dot{y}, \nonumber \\
(\dot{\delta \dot{x}}) &= - \frac{\partial^2 V}{\partial x^2}\delta x - \frac{\partial^2 V}{\partial x \partial y}\delta y, \nonumber\\
(\dot{\delta \dot{y}}) &= - \frac{\partial^2 V}{\partial y \partial x}\delta x - \frac{\partial^2 V}{\partial y^2}\delta y,
\label{variac}
\end{align}
where
\begin{align}
\frac{\partial^2 V}{\partial x^2} &= 2 y^2 \left(2x^4 - 5x^2 +1 \right) e^{-\left( x^2 + y^2 \right)}, \nonumber\\
\frac{\partial^2 V}{\partial x \partial y} &= 4 x y \left(x^2 - 1\right)\left(y^2 - 1\right) e^{-\left( x^2 + y^2 \right)}, \nonumber\\
\frac{\partial^2 V}{\partial y \partial x} &= \frac{\partial^2 V}{\partial x \partial y}, \nonumber\\
\frac{\partial^2 V}{\partial y^2} &= 2 x^2 \left(2y^4 - 5y^2 +1 \right) e^{-\left( x^2 + y^2 \right)}.
\end{align}

\begin{figure}[!t]
\begin{center}
\includegraphics[width=\hsize]{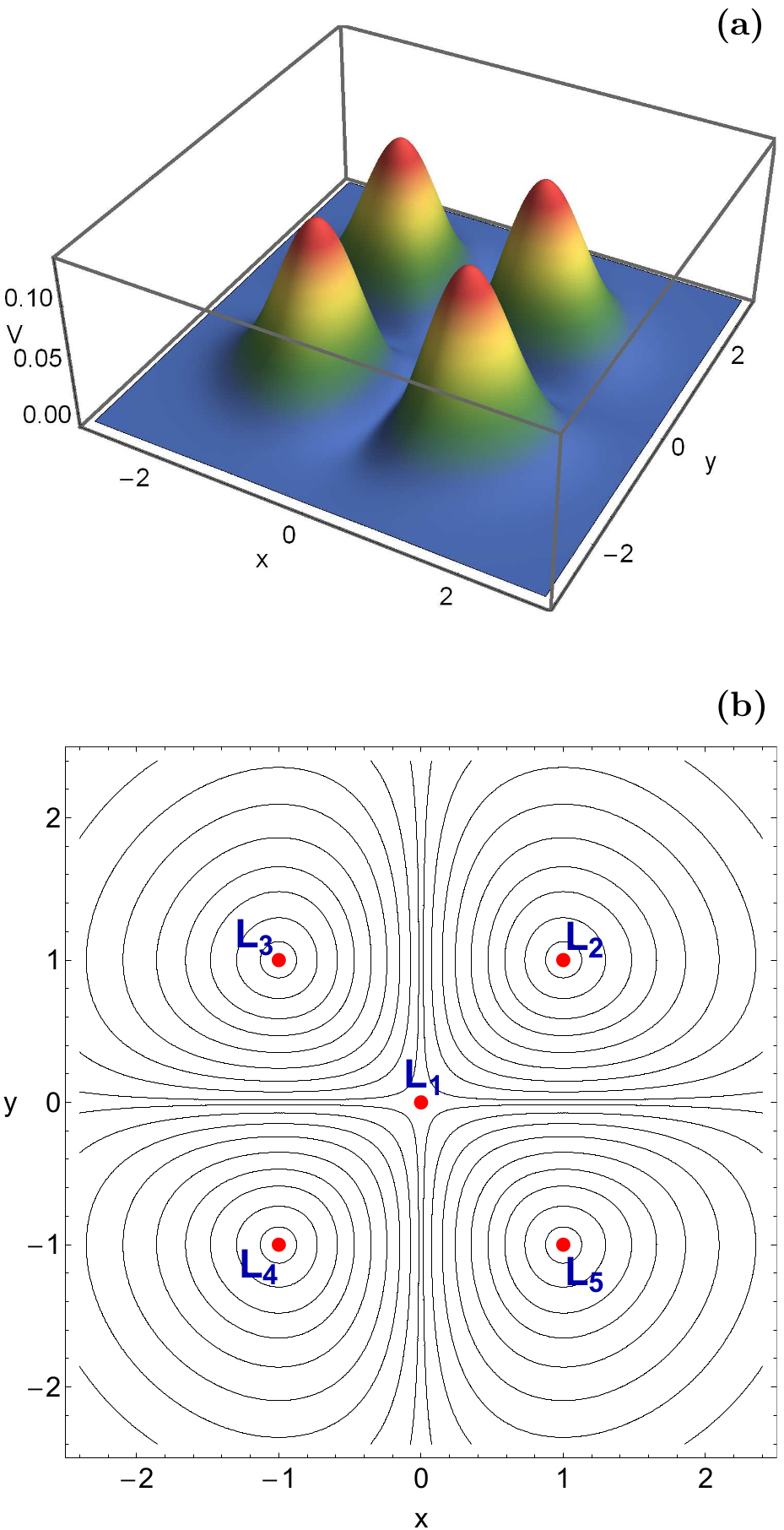}
\end{center}
\caption{(a): Surface plot of $V(x,y)$. Reddish colors indicate high values of $V(x,y)$, while blue colors correspond to low values of the potential. (b): The isoline contours of the four hill potential in the configuration $(x,y)$ space. Included are the five equilibrium points, $L_i$, $i = 1,...,5$, the positions of which are represented by red dots.}
\label{pot}
\end{figure}

The Hamiltonian to potential (\ref{phill}) reads
\begin{equation}
H(x,y,\dot{x},\dot{y}) = \frac{1}{2}\left(\dot{x}^2 + \dot{y}^2 \right) + V(x,y) = E,
\label{ham}
\end{equation}
where $\dot{x}$ and $\dot{y}$ are the velocities, while $E$ is the numerical value of the Hamiltonian, which is conserved. Thus, an orbit with a given value of energy is restricted in its motion to regions in which $E \leq V(x,y)$, while all other regions are energetically forbidden to the test particle.

The four hill potential has five equilibrium points at which
\begin{equation}
\frac{\partial V}{\partial x} = \frac{\partial V}{\partial y} = 0.
\label{sys}
\end{equation}
Four local maxima symmetrically located at $(\pm 1, \pm 1)$ (see panel (a) of Fig. \ref{pot}). The potential and its partial derivatives are identically 0 along the lines $x = 0$ and $y = 0$. The origin $(0,0)$ is the intersection of these two lines. Therefore the origin, with $E = 0$, is a degenerate extremal parabolic point. In panel (b) of Fig. \ref{pot} we present the isoline contours of the four hill potential on the configuration $(x,y)$ space. The positions of the five equilibrium points are pinpointed by red dots. At the four local maxima the value of the potential is $E_L = 1/e^{2}$, while for larger values of the energy the energetically forbidden regions disappear.

\section{Computational methods}
\label{cometh}

In order to explore the escape dynamics of the four hill potential we need to define sets of initial conditions of orbits. For this task we define for each value of the energy integral of motion dense uniform grids of initial conditions regularly distributed inside the scattering region. Our investigation takes place in both the 2D and 3D phase space in an attempt to obtain a spherical and a more complete view of the escape mechanism of the orbits. For the 2D phase space we define a grid of $1024 \times 1024$ initial conditions on the configuration $(x,y)$ plane, while for the 3D phase space a grid of $300 \times 300 \times 300$ initial conditions is defined inside the three-dimensional $(x,y,\dot{x})$ phase space.

An issue of paramount importance is the determination of the position as well as the time at which an orbit escapes. Usually in Hamiltonian systems, at every opening of the zero velocity curves (ZVCs) there is a highly unstable periodic orbit close to the line of maximum potential \cite{C79} which is called a Lyapunov orbit \cite{L49}. Such an orbit reaches the ZVC, on both sides of the opening and returns along the same path thus, connecting two opposite branches of the ZVC. The Lyapunov orbits are used for determining the escapes of the orbits. In particular, an orbit is considered as an escaping one when it intersects one of the Lyapunov orbits with velocity pointing outwards.

Usually the Lyapunov orbits and also the normally hyperbolic invariant manifolds (NHIMs), for systems with more degrees of freedom, grow out of the saddle points of the potential at the energy of escape. The Lyapunov orbits as well as the NHIMs are born normally hyperbolic. In the four hill potential the lines $x = 0$ and $y = 0$ are the escape channels and therefore there are no index-1 saddle points along these channels. This is exactly the reason why there are no usual Lyapunov orbits. For a 2-dof system the NHIMs are just the Lyapunov orbits, so this Hamiltonian system also does not contain usual NHIMs over index-1 saddle points. This is why the four hill system is pathological, because the bottom line of the four escape channels runs at a constant value of the energy. Therefore along the bottom line of the channels there are no index-1 saddle points of the potential. This is very unusual and it is not structurally stable. The essence of the pathology is to have escape channels without index-1 saddles.

In order to define escape in this dynamical system we shall apply the geometrical escape criterion adopted in \cite{SS13}. According to this criterion, an orbit is considered to escape when $x^2 + y^2 > R^2$, where $R = 10$. This allows us to correctly determine the escape of orbits with initial conditions inside a scattering region of a limit circle with radius $R$ from the center of the scattering region to the
region where the potential is negligible.

\begin{figure}[!t]
\begin{center}
\includegraphics[width=\hsize]{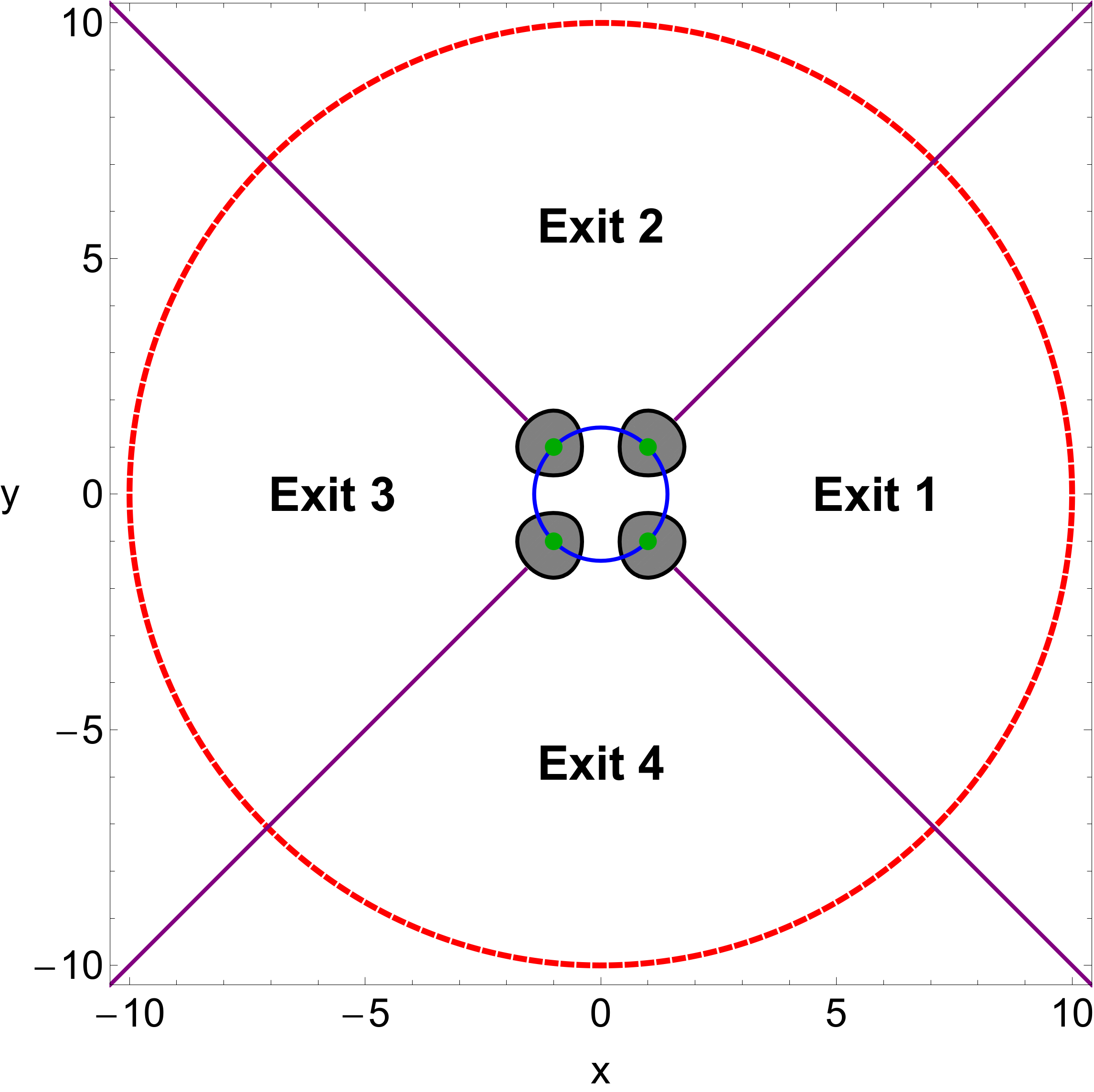}
\end{center}
\caption{The isoline contours of $V(x,y)$ for $E = 0.05$ are shown in black, while the energetically forbidden regions are plotted in gray. The positions of the four maxima are indicated by green dots. The blue circle defines the scattering region, while the outer red, dashed circle delimits the boundary between bounded and escaping motion. The purple straight lines, $y = x$ and $y = - x$, divide the configuration space into four symmetrical escape sectors.}
\label{exits}
\end{figure}

In our case, the scattering region is a circle (or a sphere) with center at the origin and radius $r_L = \sqrt{2}$ which passes through the four local maxima of the hill potential. The straight lines $y = x$ and $y = - x$, which also pass through the four maxima, divide the configuration space (and also the phase space) into four symmetrical and equal sectors. As we can see in Fig. \ref{exits} the first sector is defined for $\theta \geq 7\pi/4$ and $\theta < \pi/4$, the second sector for $\pi/4 \leq \theta < 3\pi/4$, the third sector for $3\pi/4 \leq \theta < 5\pi/4$ and the fourth sector for $5\pi/4 \leq \theta < 7\pi/4$, where the origin of the angle $\theta$ is at the center $(0,0)$.

In Hamiltonian systems the configuration as well as the phase space is divided into the escaping and non-escaping (bounded) regions. Usually, the non-escaping space is occupied by initial conditions of regular orbits forming stability islands where a third adelphic integral of motion is present. In many systems however, trapped chaotic orbits have also been observed (e.g., \cite{Z15a}). Therefore, we decided to distinguish between regular non-escaping and trapped chaotic motion.

Over the years, several chaos indicators have been developed in order to determine the character of orbits. In our case, we choose to use the Smaller ALingment Index (SALI) method. The SALI \cite{S01} has been proved a very fast, reliable and effective tool. The mathematical definition of SALI is the following
\begin{equation}
\rm SALI(t) \equiv min(d_-, d_+),
\label{sali}
\end{equation}
where $d_- \equiv \| {\vec{w_1}}(t) - {\vec{w_2}}(t) \|$ and $d_+ \equiv \| {\vec{w_1}}(t) + {\vec{w_2}}(t) \|$ are the alignments indices, while ${\vec{w_1}}(t)$ and ${\vec{w_2}}(t)$, are two deviation vectors which initially point in two random directions. For distinguishing between ordered and chaotic motion, all we have to do is to compute the SALI along a time interval $t_{\rm max}$ of numerical integration. In particular, we track simultaneously the time-evolution of the main orbit itself as well as the two deviation vectors ${\vec{w_1}}(t)$ and ${\vec{w_2}}(t)$ in order to compute the SALI.

The time-evolution of SALI strongly depends on the particular nature of the computed orbit. More precisely, if an orbit is regular the SALI exhibits small fluctuations around non zero values, while on the other hand, in the case of chaotic orbits the SALI, after a small transient period, it tends exponentially to zero approaching the limit of the accuracy of the computer $(10^{-16})$. Therefore, the particular time-evolution of the SALI allows us to distinguish fast and safely between regular and chaotic motion. Nevertheless, we have to define a specific numerical threshold value for determining the transition from order to chaos. Our previous numerical experience \cite{SABV04} suggests that if SALI $> 10^{-4}$ the orbit is ordered, while if SALI $< 10^{-8}$ the orbit is surely chaotic. On the other hand, when the final value of SALI lies in the interval $10^{-4} \leq$ SALI $\leq 10^{-8}$ we have the case of a sticky orbit\footnote{A sticky orbit is a special type of orbit which behave as a regular one for long time intervals before it exhibits its true chaotic nature.} and further numerical integration is needed so as to fully reveal the true character of the orbit.

For the numerical integration we set a maximum time equal to $10^4$ time units. Our previous experience in the subject of escaping particles indicates that usually orbits need considerable less time to find one of the exits in the equipotential surface and eventually escape from the system (obviously, the numerical integration is effectively ended when an orbit passes through one of the escape channels and intersects the limit circle). Nevertheless, we decided to use such a vast integration time just to be sure that all orbits have enough time in order to escape. Here we should clarify that orbits which do not escape after a numerical integration of $10^4$ time units are considered as non-escaping or trapped.

A double precision Bulirsch-Stoer algorithm written in standard \verb!FORTRAN 77! (e.g., \cite{PTVF92}) was used in order to forward integrate the equations of motion (\ref{eqmot}) as well as the variational equations (\ref{variac}) for all the initial conditions of the orbits. Throughout all our computations, the energy integral of motion of Eq. (\ref{ham}) was conserved better than one part in $10^{-13}$, although for most orbits it was better than one part in $10^{-14}$. All the graphical illustration of the paper has been created using the latest version 11 of Mathematica$^{\circledR}$ \cite{W03}.

\section{Escape dynamics}
\label{numres}

\begin{figure}[!t]
\begin{center}
\includegraphics[width=\hsize]{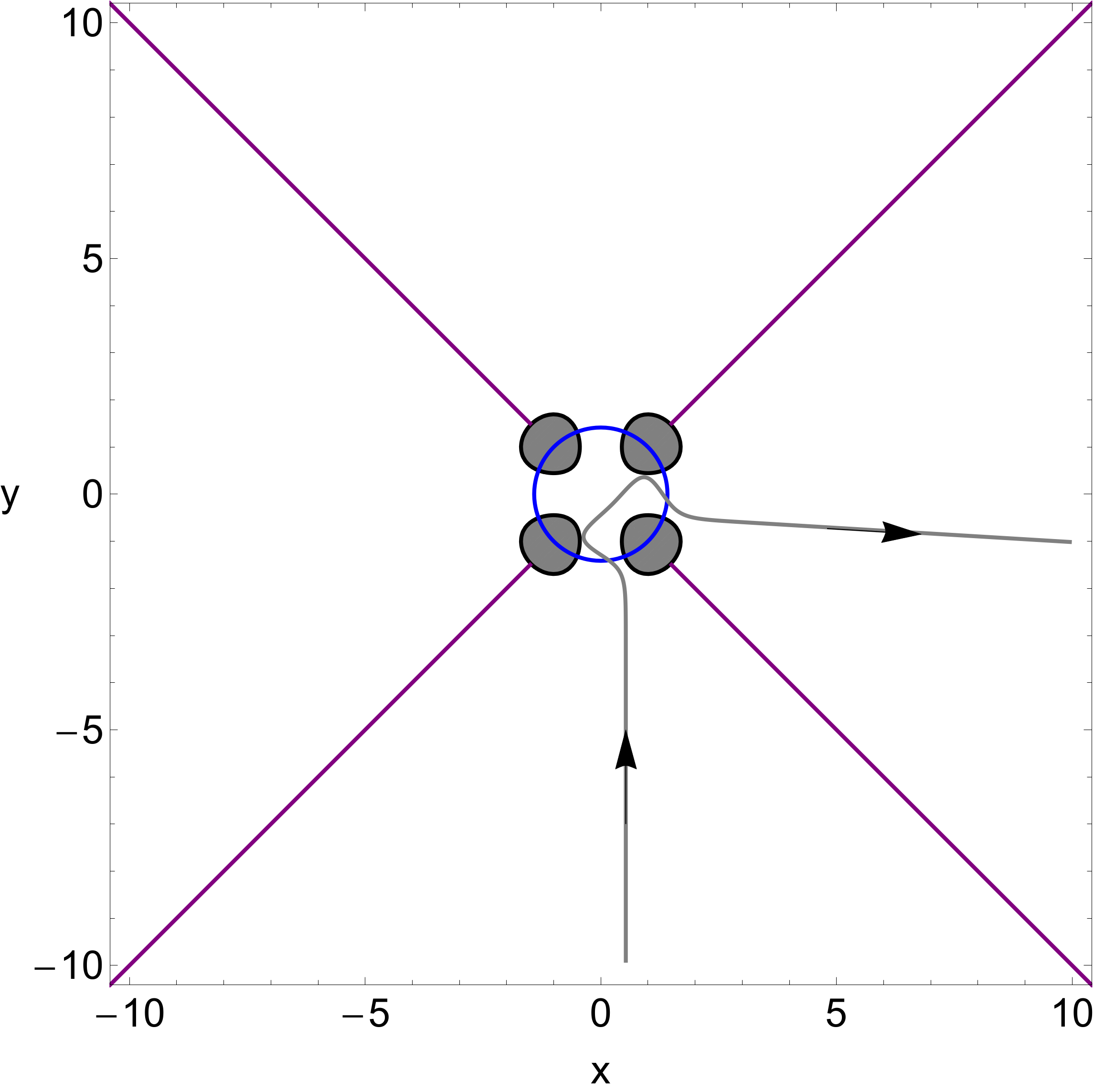}
\end{center}
\caption{A typical example of chaotic scattering for $E = 0.06$. An orbit comes from infinity, enters the central scattering region of the potential (inside the blue circle), suffers collisions between the walls of the ZVC and then being scattered thus escaping through exit channel 1. We observe that after the completion of the phenomenon of the chaotic scattering the orientation of the orbit has been changed almost by $\pi/2$.}
\label{scat}
\end{figure}

Our main target in this section will be to distinguish between escaping and non-escaping orbits for values of energy in the interval $E \in (0, E_L]$. Here it should be noted that for energy levels above $E_L$ the energetically forbidden regions disappear and therefore the concept of four escape channels has no physical meaning, since all the available phase space is available for motion. Furthermore, two important properties of the orbits will be investigated: (i) the direction or channel through which the test particles escape and (ii) the time-scale of the escapes (we shall also use the term escape period).

In our work we shall deal only with initial conditions inside the central scattering region of the potential. Of course, orbits coming from infinity may enter the central region of the potential and being scattered. In Fig. \ref{scat} we present such a typical example of scattering orbit, for $E = 0.06$.

\subsection{Results for the 2D phase space}
\label{ss1}

\begin{figure*}[!t]
\centering
\resizebox{\hsize}{!}{\includegraphics{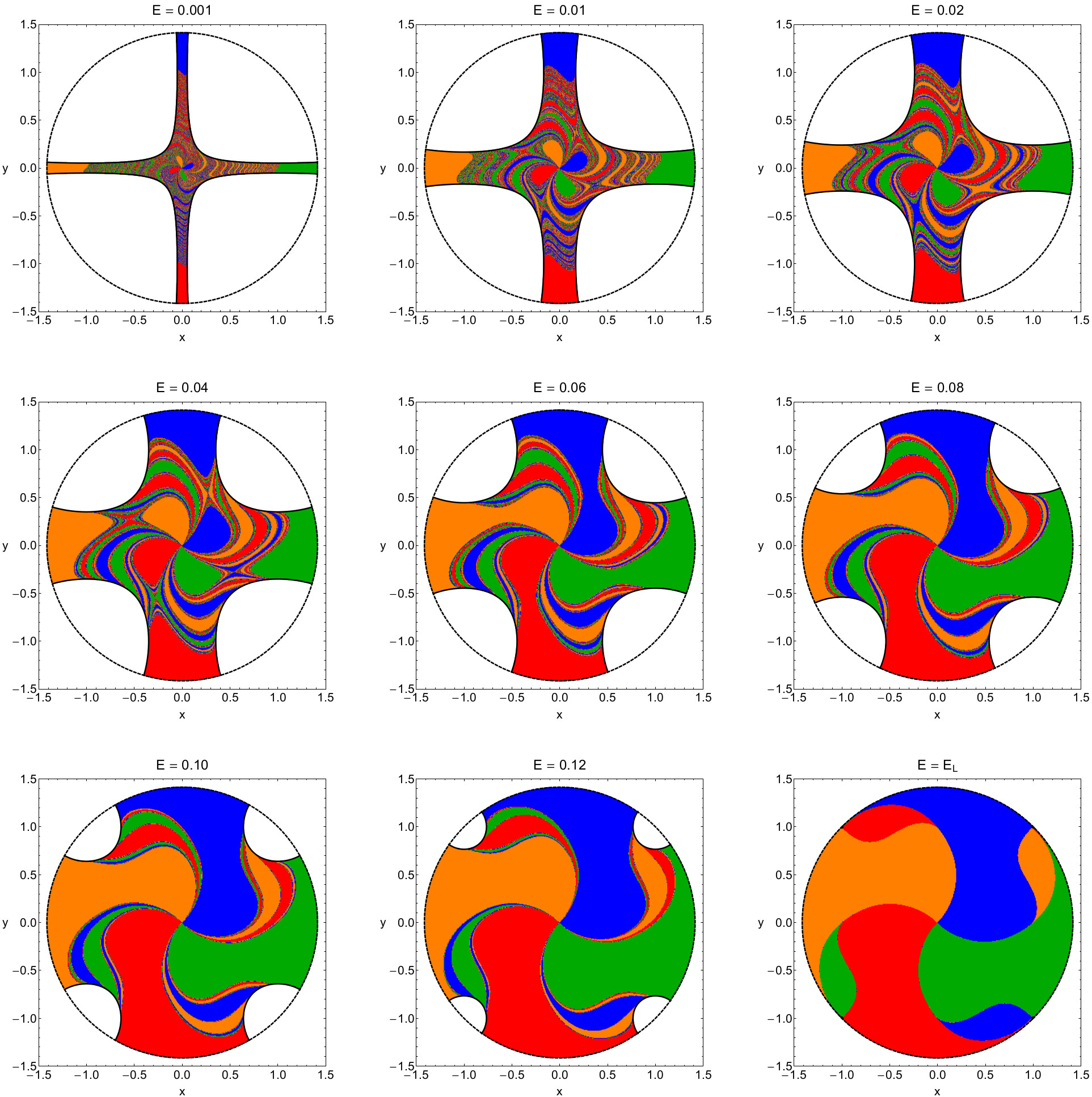}}
\caption{The orbital structure of the configuration $(x,y)$ plane for several values of the energy $E$. The color code is as follows: Escape through channel 1 (green); escape through channel 2 (blue); escape through channel 3 (orange); escape through channel 4 (red); non-escaping regular orbits (cyan); trapped chaotic orbits (yellow).}
\label{xy}
\end{figure*}

\begin{figure*}[!t]
\centering
\resizebox{\hsize}{!}{\includegraphics{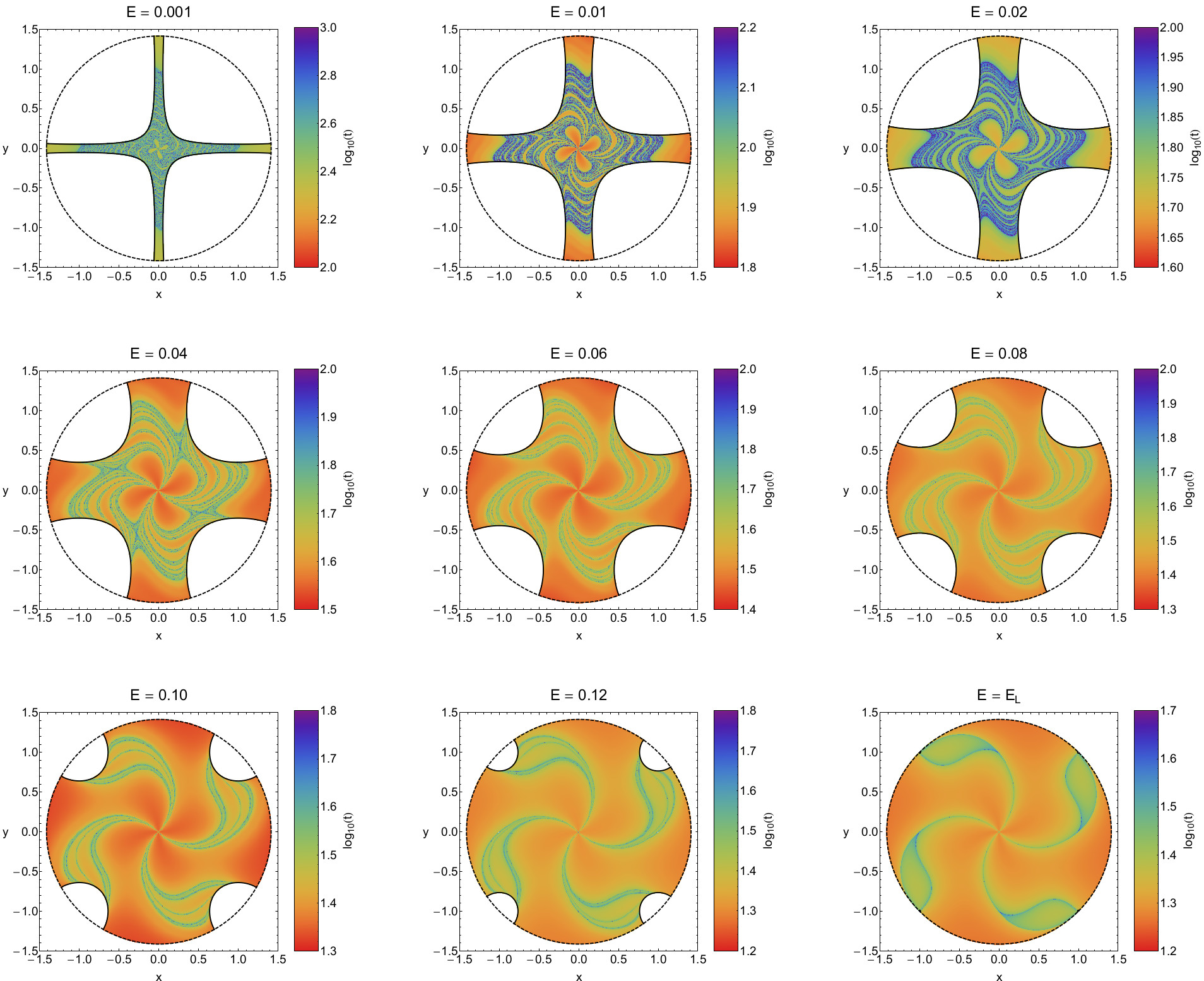}}
\caption{Distribution of the escape time $t_{\rm esc}$ of the orbits on the configuration $(x,y)$ plane. The bluer the color, the larger the escape time. Trapped chaotic and non-escaping regular orbits are shown in white. Note that the values of the colorbar are in logarithmic scale.}
\label{xyt}
\end{figure*}

Our exploration begins in the 2D phase space and particularly in the configuration $(x,y)$ space. In Fig. \ref{xy} we present the orbital structure of the $(x,y)$ plane for values of energy in the set $E = $ (0.001, 0.01, 0.02, 0.04, 0.06, 0.08, 0.10, 0.12, $E_L$). The sets of the initial conditions of the orbits are defined as follows: In polar coordinates $(r,\phi)$ the condition $\dot{r} = 0$ defines a two-dimensional surface of section, with two disjoint parts $\dot{\phi} < 0$ and $\dot{\phi} > 0$. Each of these two parts has a unique projection onto the configuration $(x,y)$ space. We choose to work on the $\dot{\phi} > 0$ part. The conditions $\dot{\phi} > 0$ and $\dot{r} = 0$, along with the existence of the energy integral of motion (\ref{ham}), suggest that for a pair of initial conditions $(x_0,y_0)$ the initial velocities in Cartesian coordinates are given by
\begin{eqnarray}
\dot{x_0} &=& - \frac{y_0}{r}\sqrt{2(E - V(x_0,y_0))}, \nonumber\\
\dot{y_0} &=& \frac{x_0}{r}\sqrt{2(E - V(x_0,y_0))},
\label{ics}
\end{eqnarray}
where $r = \sqrt{x_0^2 + y_0^2}$.

The choice of the polar coordinates can be justified as follows: the four hill potential admits a $\pi/2$ symmetry. This means that all four escape channels should be equiprobable because. In order to maintain this symmetry on the conﬁguration $(x,y)$ space we should use initial conditions expressed in polar coordinates, rather than in Cartesian coordinates. The approach of polar coordinates has been also used for the H\'{e}non-Heiles system which admits a similar symmetry of $2\pi/3$ (e.g., \cite{AVS01,AVS09}).

Each initial condition is colored according to the escape channel through which the particular orbit escapes. The cyan regions on the other hand, denote initial conditions where the test particles move in regular orbits and do not escape, while trapped chaotic orbits are indicated in yellow. The outermost solid line is the ZVC, which is defined as $V(x,y) = E$, while the black dashed circle defines the scattering region.

\begin{figure}[!t]
\begin{center}
\includegraphics[width=\hsize]{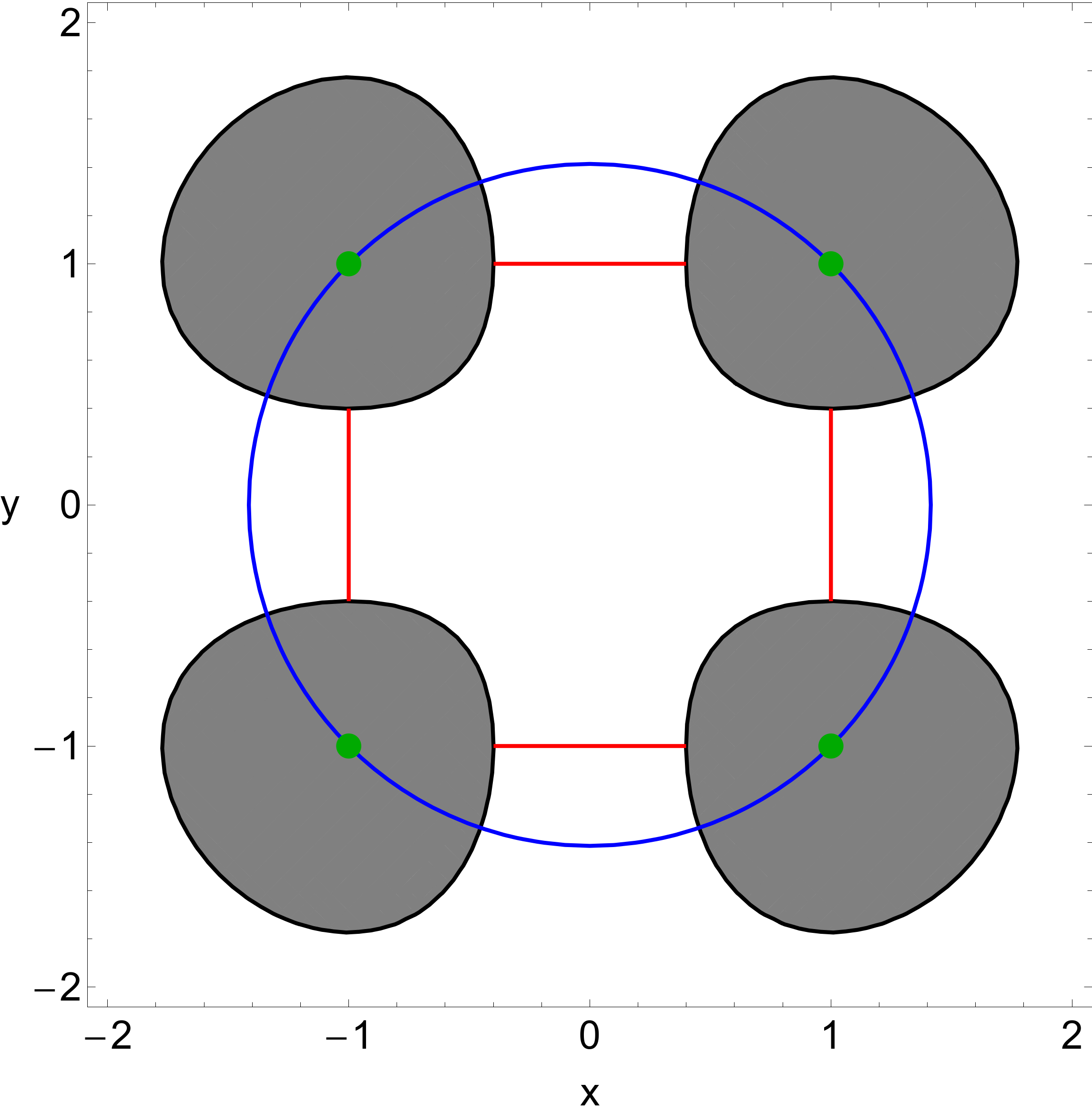}
\end{center}
\caption{The four straight-line periodic orbits (red) of the four hill potential, for $E = 0.05$. The positions of the four maxima are indicated by green dots, while the blue circle defines our scattering region. Note that the expansions of the periodic orbits inside the energetically forbidden regions (gray) pass through the four maxima, regardless the particular value of the total orbital energy.}
\label{per}
\end{figure}

It is seen that for $E = 0.001$ in the majority of the central region of the potential there is a highly sensitive dependence of the escape process on the initial conditions. Indeed, a slight change in the initial conditions makes the test particle to escape through a different channel, which is of course a classical indication of chaos. However, basins of escape are also present. By the term basin of escape, we refer to a local set of initial conditions that corresponds to a certain escape channel. These basins are mainly located either to the center of the grid forming a propeller-shape structure or to the outer parts of the scattering region. As we proceed to higher energy levels it is seen that the fractal regions reduce and the area of the basins of escape grows significantly. In particular for $E > 0.04$ the propeller-shape basins of escape merge with the corresponding ones located at the outer parts of the scattering region.

Here we would like to emphasize that when we state that an area is fractal we simply mean that it has a fractal-like geometry without conducting, at least for now, any specific calculations as in \cite{AVS09}. The fractality is strongly related with the unpredictability of a Hamiltonian system. In our case, it can be interpreted that for high enough energy levels the test particles escape very fast from the scattering region and therefore, the predictability of the dynamical system increases. Looking at Fig. \ref{xy} we see that in almost all cases (energy levels) the boundaries between the escape basins are fractal. Only for $E = E_L$ we may say that there is no numerical indication of fractal regions whatsoever\footnote{The phenomenon of almost zero fractality, at the basin boundaries, at relatively high energy levels seems to be common in several types of Hamiltonian systems. For example the same behaviour is also observed in the H\'{e}non-Heiles system but only for high enough values of the total orbital energy $(E > 2)$.}. The existence of fractal basin boundaries is a very common phenomenon observed in open Hamiltonian systems (e.g., \cite{BGOB88,dML99,dMG02,LT11,STN02,ST03,TSPT04}). However with increasing energy the basin boundaries become more more smooth which means that the degree of fractality reduces.

The distribution of the escape time $t_{\rm esc}$ of orbits on the configuration $(x,y)$ space is given in the Fig. \ref{xyt}, where light reddish colors correspond to fast escaping orbits, dark blue, purple colors indicate large escape periods, while white color denote both trapped chaotic and non-escaping regular orbits. It is observed that for $E = 0.001$ the escape periods of the majority of the orbits with initial conditions in the central region of the potential are relatively high. This however, is anticipated because in this case the width of the four escape channels is very small and therefore, the orbits should spend much time inside the ZVC until they find one of the four openings and eventually escape to infinity. As the value of the energy increases however, the escape channels become more and more wide leading to faster escaping orbits, which means that the escape period decreases rapidly. We found that the longest escape rates correspond to initial conditions near the vicinity of the fractal regions. On the other hand, the shortest escape periods have been measured for the regions without sensitive dependence on the initial conditions, that is the basins of escape.

\begin{figure}[!t]
\begin{center}
\includegraphics[width=\hsize]{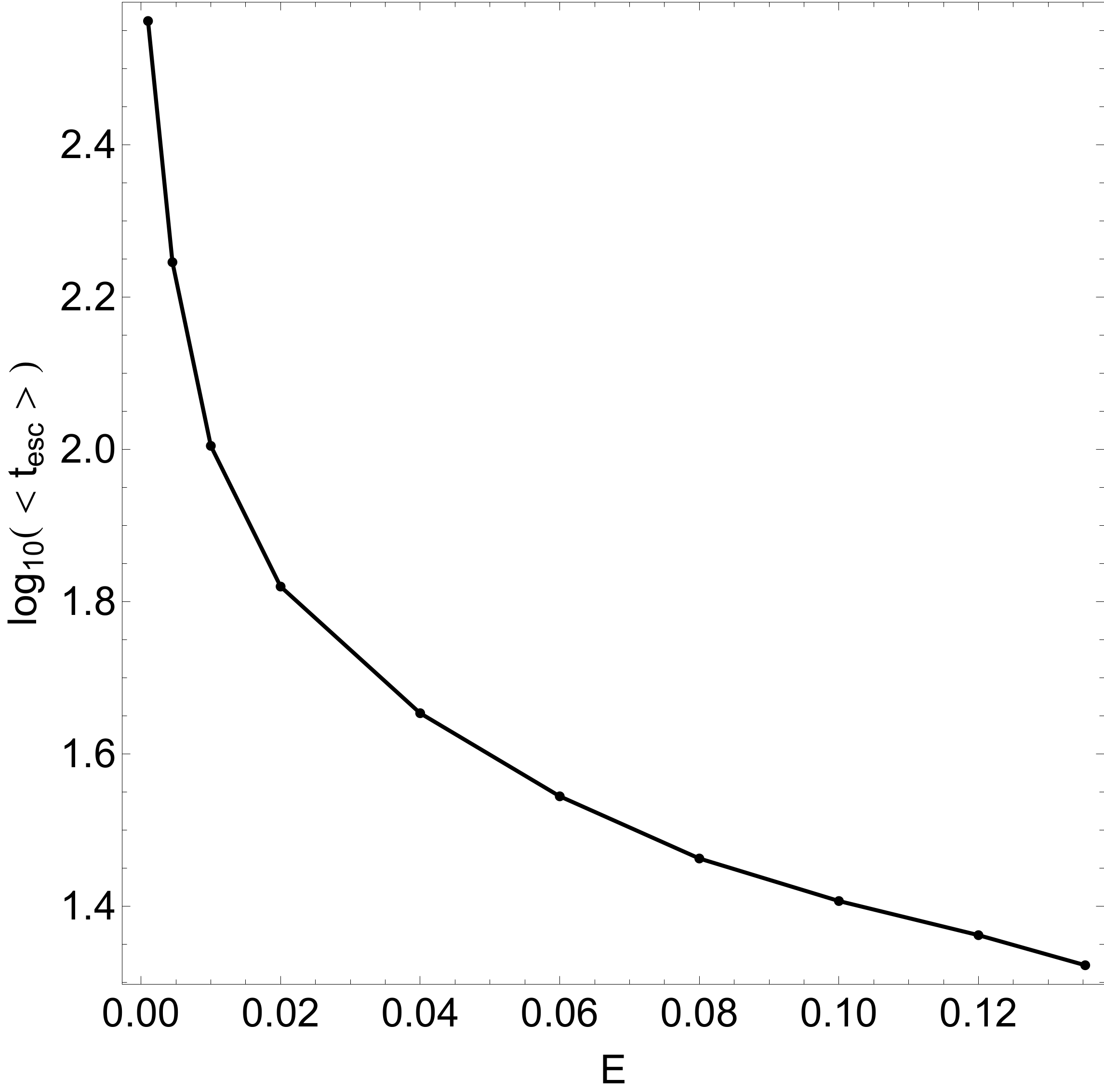}
\end{center}
\caption{Evolution of the logarithm of the average escape time of the orbits, $\log_{10}\left( < t_{\rm esc} > \right)$, as a function of the total orbital energy $E$.}
\label{tesc}
\end{figure}

\begin{figure*}[!t]
\centering
\resizebox{\hsize}{!}{\includegraphics{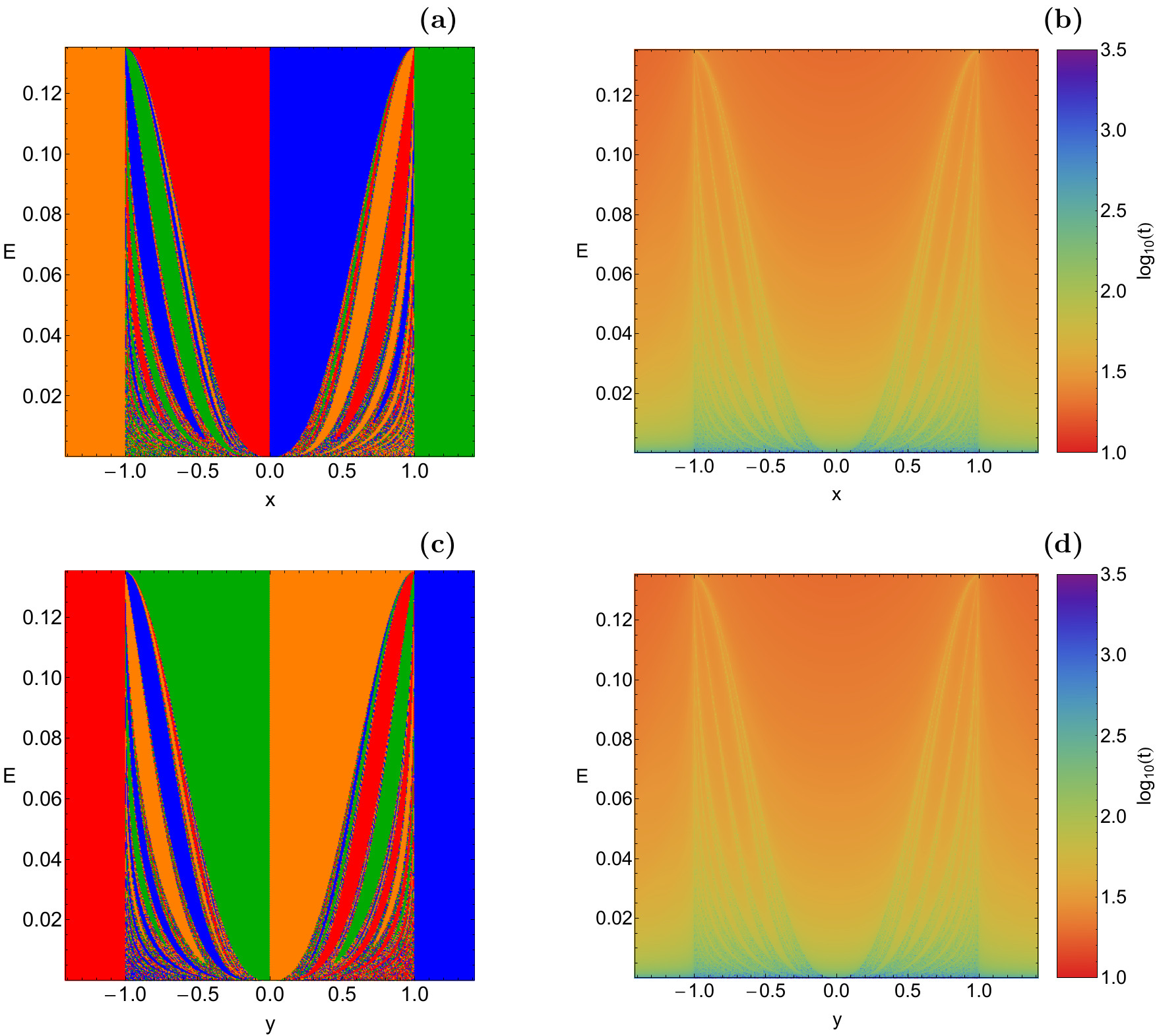}}
\caption{Orbital structure of (a-upper left): the $(x,E)$ plane and (c-lower right): the $(y,E)$ plane. The color code is the same as in Fig. \ref{xy}. Panels (b) and (d): The corresponding distribution of the escape time of the orbits of the $(x,E)$ and $(y,E)$ plane, respectively.}
\label{xyEt}
\end{figure*}

Our numerical calculations strongly suggest that in the energy range $(0,E_L]$ all tested orbits, with initial conditions inside the central scattering region, escape, sooner or later. In other words, we did not find any initial conditions corresponding neither to non-escaping regular orbits nor to trapped chaotic orbits. Taking into account that our grids of initial conditions were dense enough we may argue that in the four hill potential there is no indication of islands of bounded (regular or chaotic) motion.

Nevertheless, there must be periodic orbits oscillating between two adjacent potential hills which oscillate transverse to the escape channel. In order to check the validity of our assumption we used a numerical code for locating periodic orbits and we performed a thorough scan of the scattering region. We found that periodic orbits do exist. These periodic orbits are in fact unstable straight lines and they are located symmetrically at $x_0 = \pm 1$ and $y_0 = \pm 1$, as we can see in Fig. \ref{per} for $E = 0.05$. For the vertical periodic orbits we have $\dot{x_0} = 0$ and for the horizontal periodic orbits we have $\dot{y_0} = 0$, while in both cases the initial value of the second velocity ($\dot{y_0}$ and $\dot{x_0}$, respectively) is always obtained from the energy integral (\ref{ham}). It is interesting to note that the four straight-line periodic orbits form a square and if we expand the lines inside the energetically forbidden region they shall pass through the four maxima of the potential.

Additional numerical calculation reveal a very bizarre and pathological behavior of these straight-line periodic orbits. This is true because our computations suggest that the position of these periodic orbits remain constant, at $x_0 = \pm 1$ and $y_0 = \pm 1$, and they are completely unaffected by the change on the value of the energy. We expected the position of the periodic orbits to be a function of the total orbital energy but not in this case. These periodic orbits are generally hyperbolic. However, when the energy tends to 0 these orbits become parabolic, while for energy tending to $E_L$ they become infinitely unstable.

It would be very interesting to monitor the evolution of the average value of the escape time $< t_{\rm esc} >$ of the orbits as a function of the total orbital energy. Our results are presented in Fig. \ref{tesc}, where the values of the escape time are given in logarithmic scale. We observe that for low energy levels the average escape period of orbits is about 400 time units. However as the value of the energy increases the escape time of the orbits reduces rapidly. If we want to justify the behaviour of the escape time we should take into account the geometry of the open ZVCs. In particular, as the total orbital energy increases the four symmetrical escape channels near the maxima of the potential become more and more wide and therefore, the test particles need less and less time until they find one of the four symmetrical openings in the ZVC and escape to infinity. This geometrical feature explains why for low values of the energy orbits consume large time periods wandering inside the open ZVC until they eventually locate one of the exits and escape to infinity.

The color-coded diagrams in the configuration $(x,y)$ space, shown in Fig. \ref{xy}, provide sufficient information on the phase space mixing however, for only a fixed value of the total orbital energy and also for orbits that traverse the surface of section retrogradely. It would be very illuminating if we could scan a continuous spectrum of energy levels $E$ rather than few discrete values. In order to do this, we define a new type of a two-dimensional plane in which the value of the energy is the ordinate (see e.g., \cite{BBS08,H69}). The abscissa will be either the $x$ or the $y$ coordinate of the orbits. In particular, in the case of the $(x,E)$ plane all orbits are launched from the $x$-axis, parallel to the $y$-axis $(y = 0)$, while in the case of the $(y,E)$ plane all orbits are launched from the $y$-axis, parallel to the $x$-axis $(x = 0)$. In both cases, the initial values of the velocities $\dot{x}$ and $\dot{y}$ will be given through the Eqs. (\ref{ics}).

In panels (a) and (c) of Fig. \ref{xyEt} we present the orbital structure of the $(x,E)$ plane and $(y,E)$ plane, respectively, when $E \in (0,E_L]$, while in panels (b) and (d) of the same figure the distribution of the corresponding escape time of the orbits is depicted. It is interesting to note, that the orbital structure of both the $(x,E)$ and $(y,E)$ planes is mirror symmetrical with respect to the $y$ axis $(x = 0)$. Looking at Fig. \ref{xyEt} it becomes more than evident that for low values of the total orbital energy, $E < 0.01$, the boundaries of the several basins of escape display a large degree of fractality, while the corresponding escape time of the orbits is relatively high. However as the value of the energy increases well defined basins of escape dominate both types of planes, while the areas where the predictability of the escape process is impossible significantly decrease.

\begin{figure}[!t]
\begin{center}
\includegraphics[width=\hsize]{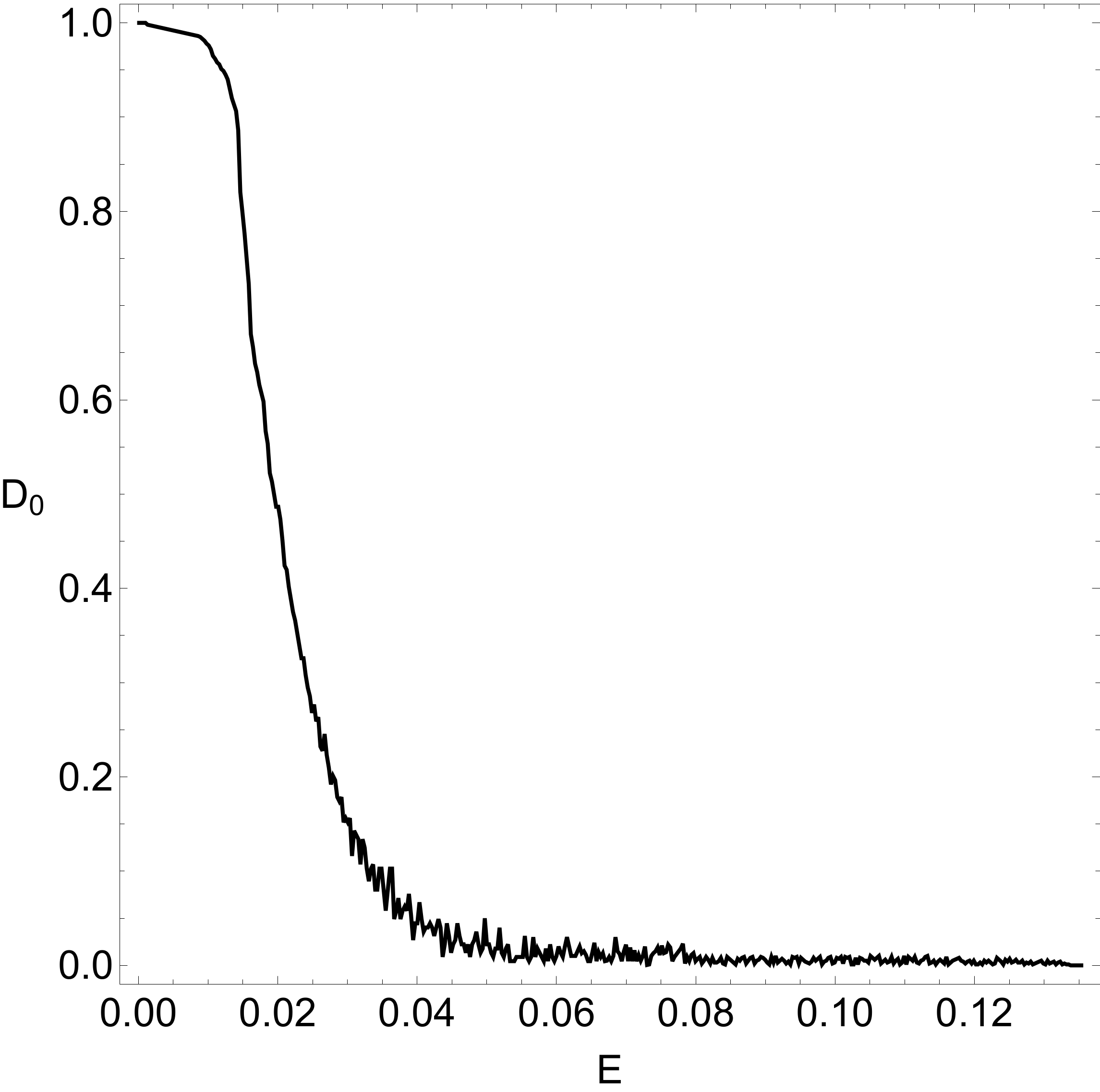}
\end{center}
\caption{Evolution of the fractal dimension $D_0$ of the $(x,E)$ plane of panel (a) of Fig. \ref{xyEt} as a function of the total orbital energy $E$. $D_0 = 1$ means total fractality, while $D_0 = 0$ implies zero fractality.}
\label{frac}
\end{figure}

So far we have discussed the fractality of the several two-dimensional planes only in a qualitative way. More precisely, we seen that the highly fractal areas are those in which we cannot predict from which escape channel each initial condition will escape. On the other hand, inside the basins of escape the degree of fractality is zero and the final state (escape channel) of the initial conditions is well known and of course predictable. At this point we shall provide a quantitative analysis regarding the degree of fractality for the $(x,E)$ plane shown earlier in panels (a) of Fig. \ref{xyEt} (due to symmetry the exact same results also apply for the $(y,E)$ plane). In order to measure the degree of fractality we have computed the uncertainty dimension \cite{O93} for a continuous spectrum of values of the total orbital energy $E$, thus following the computational method introduced in \cite{AVS01}. Obviously, this degree of fractality is completely independent of the initial conditions we use to compute it.

The evolution of the uncertainty dimension $D_0$ of the $(x,E)$ plane, as a function of the total orbital energy $E$, is shown in Fig. \ref{frac}. The computation of the uncertainty dimension was done for only a ``1D slice'' of initial conditions of Fig. \ref{xyEt}, and for that reason $D_0 \in [0,1]$. It is interesting to note that in the energy interval $0 < E < 0.008$ the uncertainty dimension tends to one. This means that for these energy levels there is a total fractalization of the $(x,E)$ plane and therefore the chaotic set becomes ``dense" in the limit. As the value of the total orbital energy increases however the degree of fractality is substantially being reduced. Being more precisely, for $E > 0.06$ the uncertainty dimension fluctuates at extremely low values $(D_0 < 0.05)$, while for $E > 0.134$ completely vanishes. This behavior explains, in a way, the phenomenon we observed earlier in the last panel of Fig. \ref{xy}, where for $E = E_L$ all basin boundaries were found to be extremely smooth without any indication of fractal regions.

At this point we would like to emphasize that there are many methods for computing the predictability of a dynamical system. Very recently a new tool, called ``basin entropy", has been developed for analyzing the uncertainty in dynamical systems \cite{DWG16}. This new qualitative method describes the notion of fractality and unpredictability in the context of basins of attraction or basins of escape. The basin entropy provides a excellent qualitative information regarding the fractality of the escape basins, so it would be very illuminating if we could determine how the basin entropy evolves as a function of the total orbital energy. In a future work we shall use this new tool in order to investigate, in several types of Hamiltonian systems, how the unpredictability is significantly being reduced as the energy increases. Furthermore, it would be very informative to compare the corresponding results derived from both the basin entropy and the uncertainty dimension.

\begin{figure*}[!t]
\resizebox{\hsize}{!}{\includegraphics{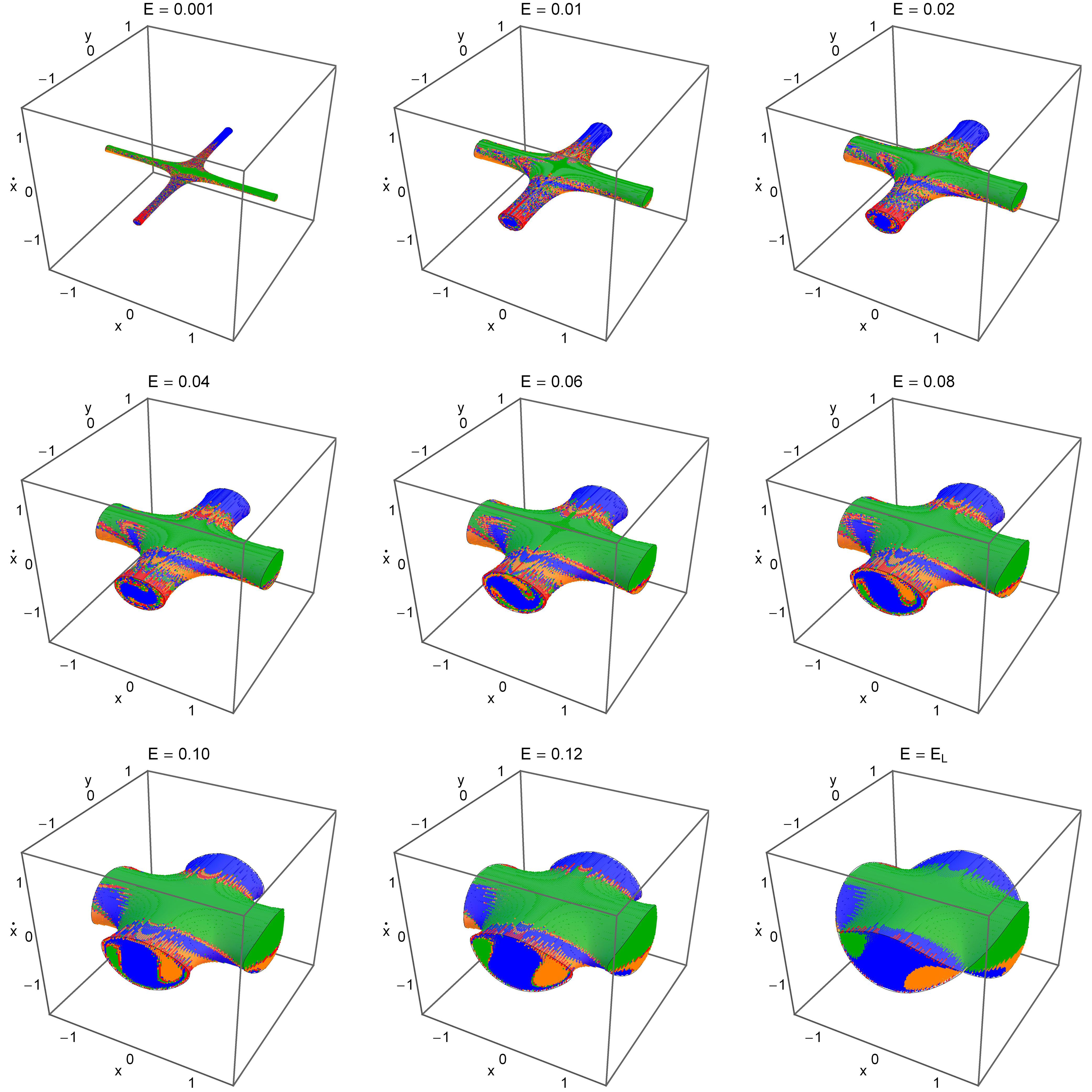}}
\caption{Orbital structure of three dimensional distributions of initial conditions of orbits in the 3D $(x,y,\dot{x})$ subspace for several values of the energy $E$. The color code is the same as in Fig. \ref{xy}.}
\label{3d}
\end{figure*}

\begin{figure*}[!t]
\resizebox{\hsize}{!}{\includegraphics{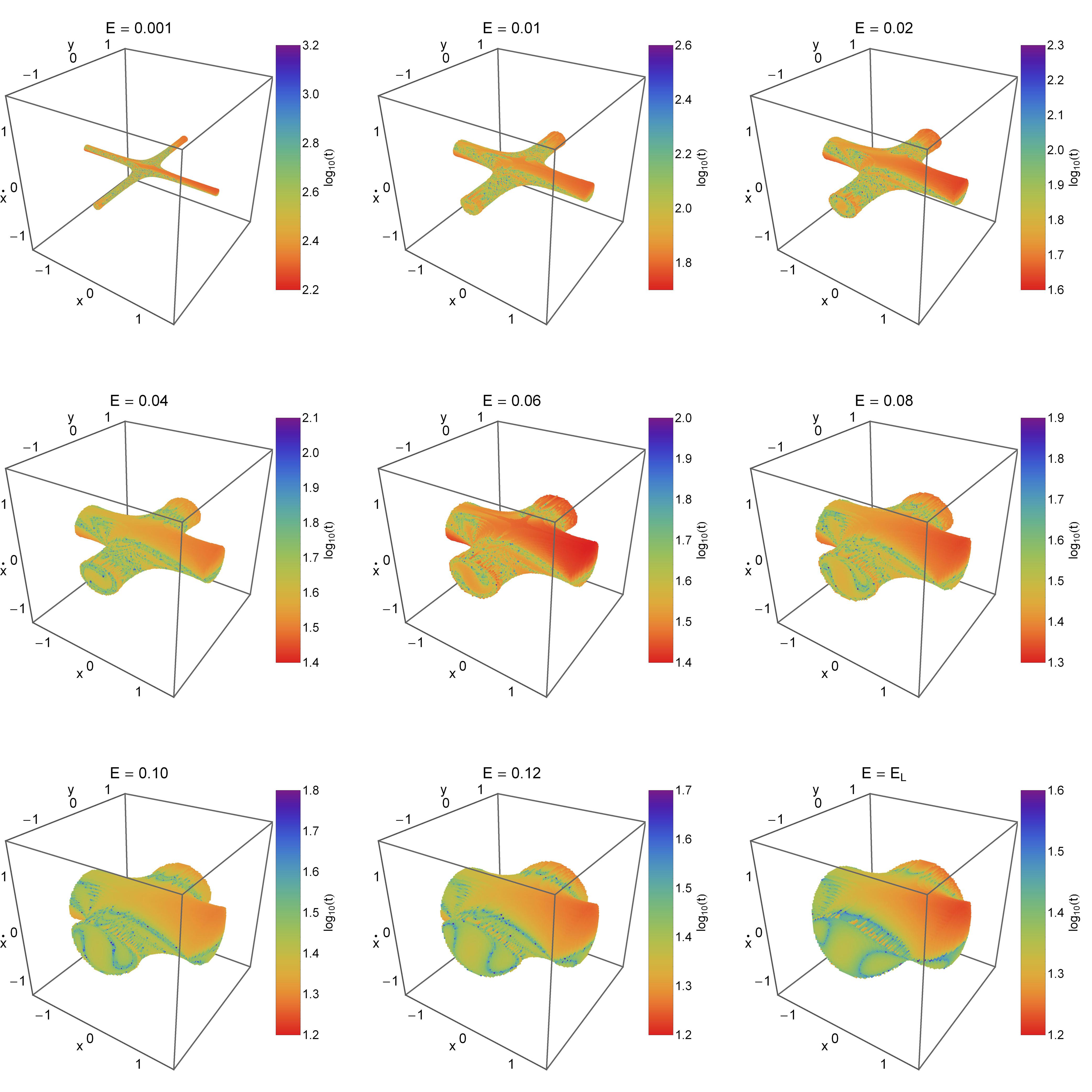}}
\caption{Distribution of the corresponding escape time of the orbits with initial conditions inside the 3D $(x,y,\dot{x})$ subspace for the values of energy presented in Fig. \ref{3d}. The bluer the color, the larger the escape time. Initial conditions of bounded orbits are shown in transparent color.}
\label{3dt}
\end{figure*}

\begin{figure*}[!t]
\resizebox{\hsize}{!}{\includegraphics{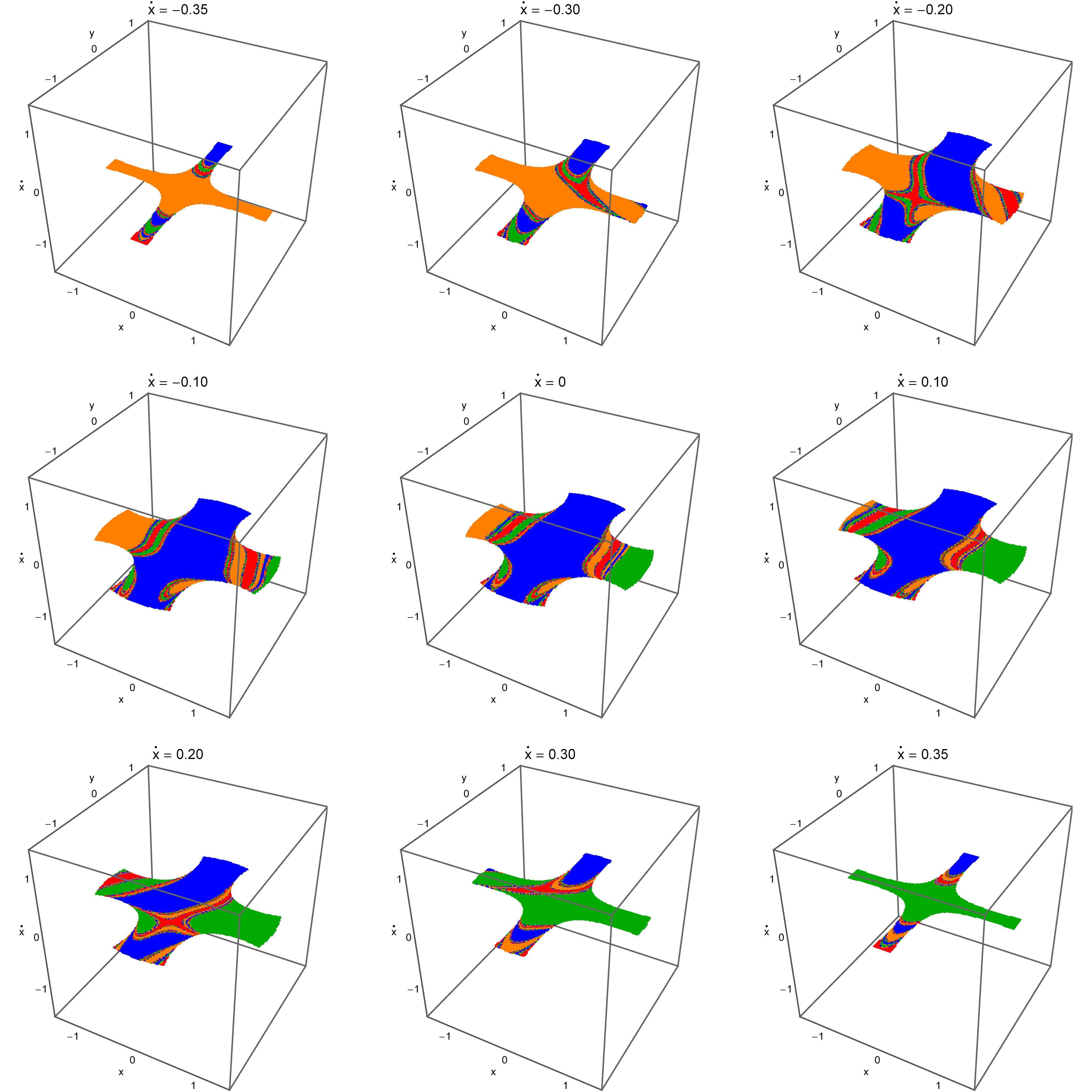}}
\caption{A tomographic view of the solid grid of Fig. \ref{3d} when $E = 0.06$ showing slices on the configuration $(x, y)$ plane when $\dot{x} = (-0.35, -0.30, -0.20, -0.10, 0, 0.10, 0.20, 0.30, 0.35)$.}
\label{tomo}
\end{figure*}

\subsection{Results for the 3D phase space}
\label{ss2}

In the previous subsection we investigated the escape dynamics of orbits using two dimensional grids of initial conditions in several types of planes (or in other words, in several 2D subspaces of the whole 4D\footnote{The dimension of the complete phase space of a Hamiltonian system with $N$ degrees of freedom is $2N$. This phase space is foliated into $2N - 1$ dimensional invariant leaves corresponding to the numerical values of the Hamiltonian. However, the dimension of the entire phase space always remains $2N$, irrelevantly of any possible invariant foliation.} phase space). In this subsection we will expand our numerical exploration using three dimensional distributions of initial conditions of orbits. Being more precise, for a particular value of the energy we define inside the corresponding zero velocity surface uniform grids of initial conditions $(x_0,y_0,\dot{x_0})$, while the initial value of $\dot{y} > 0$ is always obtained from the energy integral of motion (\ref{ham}).

In Fig. \ref{3d} we present the orbital structure of the three dimensional distributions of initial conditions of orbits in the $(x,y,\dot{x})$ subspace for the same set of values of the energy. The color code is the same as in Fig. \ref{xy}. At this point, we should emphasize that in this case the distribution of the initial conditions of the orbits cannot illustrate any longer the $\pi/2$ symmetry of the system, which is only visible in polar coordinates and only in the two-dimensional $(x,y)$ plane. This means that the four escape channels of the four hill potential are not equiprobable. However, it should be clarified that the fact that the escape channels are no longer equiprobable is just a computational effect and a numerical artifact of the particular choice of the initial conditions\footnote{The same phenomenon of the loss of the symmetry of a dynamical system due to the particular choice of the initial conditions of the orbits also applies for other types of Hamiltonian systems. For example, in the H\'{e}non-Heiles system the $2\pi/3$ symmetry is observable only in the configuration $(x,y)$ space and only by using polar coordinates. On the other hand, for initial conditions of orbits in the $(y,\dot{y})$ phase space the three escape channels are no longer equiprobable (see e.g., the series of papers on the H\'{e}non-Heiles system by Sanju\'{a}n and collaborators).}. As far as we know, there is no any choice of initial conditions that would maintain the $\pi/2$ symmetry of the system in the $(x,y,\dot{x})$ subspace.

The corresponding distributions of the escape time of the orbits are given in Fig. \ref{3dt}, where for the initial conditions of the bounded orbits we have used transparent white color in order to be able to inspect, in a way, the interior of the solids.

Once more, our numerical results indicate that all examined initial conditions correspond to escaping orbits without any trace of bounded regular or trapped chaotic motion. Perhaps there are also three-dimensional isolated unstable periodic orbits similar to the straight-line periodic orbits we found in the configuration $(x,y)$ space. However it is the author's opinion that the existence or not of these periodic orbits is out of the scope of the present article.

It is seen in Fig. \ref{3d} that the grids of the initial conditions of the orbits are in fact three dimensional solids and therefore only their outer surface is visible. However, a tomographic-style approach can be used in order to penetrate and examine the interior region of the solids (e.g., \cite{Z14a}). According to this method, we can plot two dimensional slices of the solid grid by defying specific levels to a primary plane (i.e., the $(x,y)$ plane). Fig. \ref{tomo} shows the evolution of the structure of the configuration $(x,y)$ plane when $\dot{x}$ = (-0.35, -0.30, -0.20, -0.10, 0, 0.10, 0.20, 0.30, 0.35) for $E = 0.06$. We see that the structure evolves rapidly and non-uniformly (constant interplay between escape basins and fractal structures) thus implying that the escape process in this dynamical system is, by all means, a very complex and fascinating procedure. More precisely, we observe that for low values of the velocity $\dot{x}$ escaping orbits through channel 3 dominate, around $\dot{x} = 0$ channel 2 is the dominant one, while for high values of $\dot{x}$ channel 1 seems to be more preferable. Similar tomographic plots can be obtained also for the other values of the energy presented in Fig. \ref{3d}. Our computations indicate that at least a portion of the symmetry of the system is retained in the 3D subspace because the amount of orbits that escape through exit channels 1 and 3 were found to be exactly the same.

Taking into account that the escape channels are no longer equiprobable we could present the evolution of the percentages of all types of orbits as a function of the value of the total orbital energy $E$. However we feel that this could be confusing or even misleading regarding the physics behind the problem. Once more we have to emphasize that the $\pi/2$ symmetry is destroyed due to the particular choice of the initial conditions of the orbits and not because of some internal property of the system.

Before closing this section we would like to mention that our numerical computations suggest that the average escape time of the orbits, with initial condition in the three-dimensional $(x,y,\dot{x})$ space, exhibits a similar evolution to that discussed earlier in Fig. \ref{tesc} for the distribution of orbits on the two-dimensional configuration $(x,y)$ plane.

\section{Discussion}
\label{disc}

The numerical exploration of the escape dynamics of the four hill potential was the aim of this work. For this purpose we investigated the orbital structure in many types of two-dimensional planes and for several values of the total orbital energy $E$. We also proceeded one step further by classifying initial conditions of orbits in a three-dimensional (3D) subspace of the whole four-dimensional (4D) phase space. We managed to distinguish between escaping orbits and we also located the basins of escape leading to different exit channels, finding correlations with the corresponding escape time of the orbits. Among the escaping orbits we separated between those escaping fast or late from the system. Our extensive numerical calculations strongly suggest that the overall escape process of the four hill potential is very dependent on the value of the total orbital energy.

The four hill potential is a Hamiltonian system with many strange properties, which are listed below, with respect to usual and generic dynamical systems such as the H\'{e}non-Heiles system
\begin{itemize}
  \item The energy of escape has no meaning since escape motion is possible in the entire range of the energy $(E \in [0,\infty)$. In other words, there are no energy levels for which only bounded (regular or chaotic) motion is possible.
  \item There are no index-1 saddle points and therefore there are neither usual Lyapunov orbits nor usually normally hyperbolic invariant manifolds (NHIMs) over the index-1 saddle points.
  \item The unstable periodic orbits, which oscillate between the walls of the zero velocity curves, are straight (horizontal and vertical) lines in the configuration $(x,y)$ space.
  \item The positions of the unstable periodic orbits are fixed and completely unaffected by the change on the value of the total orbital energy.
\end{itemize}
All the above-mentioned peculiar properties strongly indicate how pathological is the four hill potential. This is exactly the reason of why we decided to investigate its orbital dynamics.

For the numerical integration of the sets of the initial conditions of orbits in each type of grid (2D and 3D), we needed between about 15 minutes and 22 hours of CPU time on a Quad-Core i7 2.4 GHz PC, depending of course on the escape rates of orbits in each case. For each initial condition the maximum time of the numerical integration was set to be equal to $10^4$ time units however, when a test particle escaped the numerical integration was effectively ended and proceeded to the next available initial condition.

We hope that the present analysis and the corresponding numerical results of the four hill potential to be useful in this active field of nonlinear dynamics by shedding some light to the complicated escape mechanism of orbits. Since our results are encouraging, it is in our future plans to expand our investigation and study the escape properties of orbits in other more complicated types of open Hamiltonian systems (e.g., binary stellar systems, spiral galaxies, etc).

\section*{Acknowledgments}

I would like to gratefully and sincerely thank Dr. Christof Jung for all the illuminating discussions during this research. My warmest thanks also go to the two anonymous referees for the careful reading of the manuscript as well as for all the apt suggestions and comments which allowed us to improve both the quality and the clarity of the paper.

\section*{Compliance with Ethical Standards}

\begin{itemize}
  \item Funding: The author states that he has not received any research grants.
  \item Conflict of interest: The author declares that he has no conflict of interest.
\end{itemize}

\end{document}